\documentclass[fleqn,twoside]{article}
\usepackage{espcrc2}
\usepackage[dvips]{graphicx}
\def\Journal#1#2#3#4{{#1} {\bf #2}, #3 (#4)}

\def\be{\begin{equation}}
\def\ee{\end{equation}}
\def\bea{\begin{eqnarray}}
\def\eea{\end{eqnarray}}

\title{Measurement of the $\eta$ mass at KLOE.}
\author{
F.Ambrosino, A.Antonelli, M.Antonelli, F. Archilli, C.Bacci, P.Beltrame, G.Bencivenni, S.Bertolucci, C.Bini, C.Bloise, S.Bocchetta, V.Bocci, F.Bossi, P.Branchini, R.Caloi, P.Campana, G.Capon, T.Capussela, F.Ceradini, S.Chi, G.Chiefari, P.Ciambrone, E.De Lucia, A.De Santis, P.De Simone, G.De Zorzi, A.Denig, A.Di Domenico, C.Di Donato, S.Di Falco, \makebox{B.Di Micco}$^0$, A.Doria, M.Dreucci, G.Felici, A.Ferrari, M.L.Ferrer, G.Finocchiaro, S.Fiore, C.Forti, P.Franzini, C.Gatti, P.Gauzzi, S.Giovannella, E.Gorini, E.Graziani, M.Incagli, W.Kluge, V.Kulikov, F.Lacava, G.Lanfranchi, J.Lee-Franzini, D.Leone, M.Martini, P.Massarotti, W.Mei, S.Meola, S.Miscetti, M.Moulson, S.M\"uller, F.Murtas, M.Napolitano, F.Nguyen, M.Palutan, E.Pasqualucci, A.Passeri, V.Patera, F.Perfetto, M.Primavera, P.Santangelo, G.Saracino, B.Sciascia, A.Sciubba, F.Scuri, I.Sfiligoi, T.Spadaro, M.Testa, L.Tortora, P.Valente, B.Valeriani, G.Venanzoni, R.Versaci, G.Xu.}

\begin{document}
\begin{abstract}
An integrated luminosity of 410 pb$^{-1}$,
corresponding to $\sim$ 17 million of $\eta$ events,  has been analyzed to measure the $\eta$ mass using the decay
$\eta \to \gamma \gamma$.  The measurement is insensitive to the calorimeter energy calibration
and the systematic error on the measurement is dominated by the uniformity of the detector response. 
As a cross check of the method  the $\pi^0$ mass from the decay $\phi \to \pi^0 \gamma, \pi^0 \to \gamma \gamma$ has been measured and it is in agreement with the most accurate previous determinations. \\
The result obtained is $m_{\eta} = 547.873 \pm 0.007_{stat.} \pm 0.031_{syst.}$ MeV, that  is today  best measurement of the $\eta$ mass.
\end{abstract}

\maketitle

\footnotetext{\makebox{Corresponding author:} \\ \makebox{B. Di Micco}, \makebox{\emph{dimicco@fis.uniroma3.it}} - Universit\`a degli Studi di Roma Tre, via della Vasca Navale, 84 - 00146 Roma, Italy.}
\section{Introduction}
The KLOE experiment is performed at the Frascati $\phi$ factory DA$\Phi$NE\cite{dafne}.
DA$\Phi$NE is a high luminosity $e^+$,$e^-$ collider working at $\sqrt{s} \sim 1020$ MeV, corresponding to the  $\phi$ meson mass. 
In the whole period of data taking ($2001 - 2006$) KLOE has collected an integrated luminosity of 2.5 fb$^{-1}$, corresponding to about 8 billions of $\phi$ produced and
$~ 100$ millions of $\eta$ mesons through the electromagnetic decay $\phi \to \eta \gamma$.

The KLOE detector consists of a large cylindrical drift chamber
\cite{K-dc}, DC, surrounded by a lead/scintillating-fiber sampling
calorimeter \cite{K-emc}, EMC, both immersed in a solenoidal
magnetic field of 0.52 T with the axis parallel to the beams, $z$ in the following.
 The DC momentum 
resolution for charged particles is $\delta p_\perp/p_\perp$=0.4\%. 
 The calorimeter
is divided into a barrel and two endcaps, and covers 98$\%$ of the 
total
solid angle. Photon energies and arrival times are measured with
resolutions $\sigma_{E}/E = 0.057/{\sqrt{E \ ({\rm GeV})}}$  and
$\sigma_{t} = 54 \ {\rm ps} /{ \sqrt{E \ ({\rm GeV})}} \oplus 50 \
{\rm ps}$, respectively. Photon-shower centroid positions are measured 
with an accuracy of $\sigma=1\ {\rm cm}/\sqrt{E\ ({\rm GeV})}$ along the 
fibers, and 1 cm in the transverse direction.
A photon is defined as a cluster of energy deposits in the calorimeter 
elements that is not associated to a charged particle. We require
the distance between the cluster centroid and the nearest entry point of 
extrapolated tracks be greater than 3$\times \sigma(z,\phi)$, where $\phi$ 
is the azimuthal angle.

The trigger \cite{K-trigger} uses information from both the
calorimeter and the drift chamber. The EMC trigger requires two
local energy deposits above threshold ($E\!>\!50$ MeV in the barrel,
$E\!>\!150$ MeV in the endcaps).  The
trigger has a large time spread  with respect to the time distance
between consecutive beam crossings. It is however synchronized
with the machine radio frequency divided by four, $T_{\rm sync}$=10.85 
ns, with an accuracy of 50 ps. For the  2001-2002 data
taking, the bunch crossing period was $T$=5.43 ns. The time ($T_0$) of
the bunch crossing producing an event is determined offline during
event reconstruction.

\section{Measurement of the $\eta$ mass.}
The value of the $\eta$ mass has been recently measured with high precision by two collaborations NA48\cite{NA48} ($m_{\eta} = 547.843 \pm 0.030 \pm 0.041$ MeV/c$^2$) 
and GEM\cite{GEM} ($m_{\eta} = 547.311 \pm 0.028 \pm 0.032$ MeV/c$^2$) using different techniques and production reactions. The two measurements differ by more than eight standard 
deviations from each other. The GEM measurement is in agreement with the older ones\cite{PDG} while the NA48 measurement is higher. 
For this reason it is interesting to provide a further measurement of comparable precision in order to clarify the experimental situation.\section{Measurement method.} 
We measure the mass studying the decay $\phi \to \eta \gamma, \eta \to \gamma \gamma$. A kinematic fit is performed imposing the 4 constraints 
given by the energy-momentum conservation. 
Sing there are three photons and there are 4 constraints, the fit overconstrains the energies of the photons that are, practically, determined by the position of the 
clusters in the calorimeter. The inputs of the fit are the energy, the position and the time of the calorimeter clusters,
the mean position of the $e^+ e^-$ interaction point, the total four-momentum of the colliding $e^+ e^-$ pair. Each of these variables is
determined run by run using $e^+ e^- \to e^+ e^-$  events ( almost 90000 events for each run, allowing a very precise determination of the relevant parameters).  
\section{Selection.} 
\begin{figure}
\includegraphics[width=0.5\textwidth]{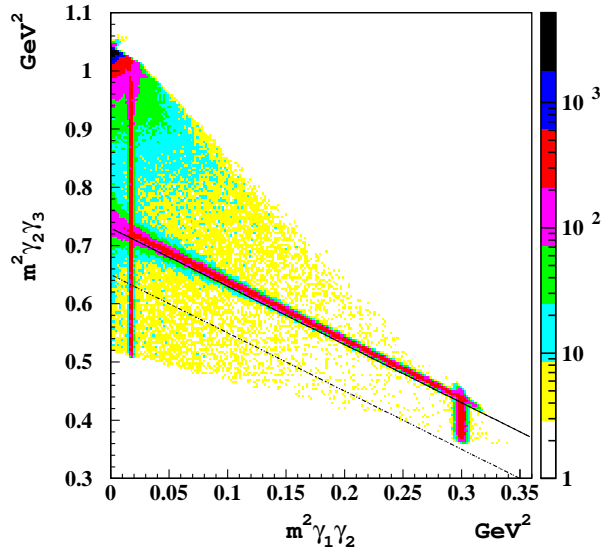}
\caption{Dalitz plot distribution of the selected 3$\gamma$ events. The photons are sorted according their energies $E_1 < E_2 < E_3$.} \label{fig:dalitz}
\end{figure}

The $\phi \to \eta \gamma$ events are selected by requiring at least three energy deposits in the barrel with a polar angle $\theta_{\gamma}: 50^{\circ} < \theta_{\gamma} < 130^{\circ}$, not associated to a charged track. A kinematic fit imposing energy-momentum conservation and time of flight of photons equal to the velocity of light is done for all 3 $\gamma$'s combination of N detected photons. The combination with the lowest $\chi^2$ is chosen as a candidate event if $\chi^2 < 35$.  The events surviving the cuts are shown in fig.\ref{fig:dalitz}, where the
Dalitz plot  is shown. Three bands are clearly visible. The band at low $m^2_{\gamma \gamma}$ is given by the $\phi \to \pi^0 \gamma, \pi^0 \to \gamma \gamma$, while the other two bands are $\phi \to \eta \gamma, \eta \to \gamma \gamma$ events.

\begin{figure} 
\includegraphics[width=0.5\textwidth]{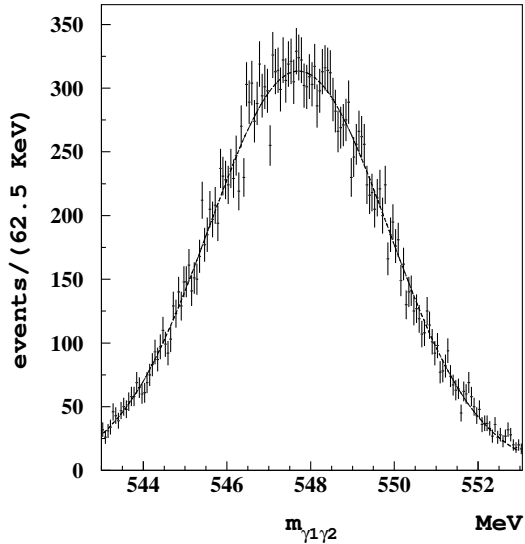}
\caption{Invariant mass distribution of the $\gamma_1 \gamma_2$ pair. The fitted function is the continuous line.} \label{fig:etafit}
\end{figure}

 With the  cut shown in the Dalitz plot we select a pure sample of $\eta, \pi^0 \to \gamma \gamma$ events. The resulting $m_{\gamma \gamma}$ spectrum (fig.\ref{fig:etafit}) can be fitted well  with a single gaussian of $\sigma \sim 2.1 MeV/c^2$ . 

In order to determine the systematic error we have evaluated the uncertainities on all the quantities used in the kinematic fit
and the effect on the fitted value. \\
A sample of $e^+ e^- \to \pi^+ \pi^- \gamma$ events has been used to check the mean position of the interaction point, the energy response of the calorimeter and  the alignment of the calorimeter with the Drift Chamber. The
mean position of the interaction point, determined run by run using the $e^+ e^- \to e^+ e^-$ events, has been compared with the reconstructed $\pi^+ \pi^-$ vertex. The difference between this two values has been computed run by run and the spread of the distribution is used for the systematic error on the determination of the I.P mean position. 
Since the I.P position is determined with the tracks reconstructed in the Drift Chamber, small displacement between the DC and the calorimeter can affect position measurement of this point in the calorimeter reference frame. To check disalignment between the calorimeter and the Drift Chamber, the $\pi^-$ and $\pi^+$ tracks of the $\pi^+,\pi^- \gamma$ events were extrapolated to the calorimeter and the closest approach point to the cluster centroid was determined. The difference in the position $\vec{x}_{clu}-\vec{x}_{cst}$ were determined and the spread of these values are taken as systematic error on DC-Calo allignment. A small correction of 1.1 mm in the Calorimeter position along the $y$ direction, the vertical, and 2 mm along the $z$ direction (the direction of the beam axis) was applied. 

\begin{figure}
\includegraphics[width=0.5\textwidth]{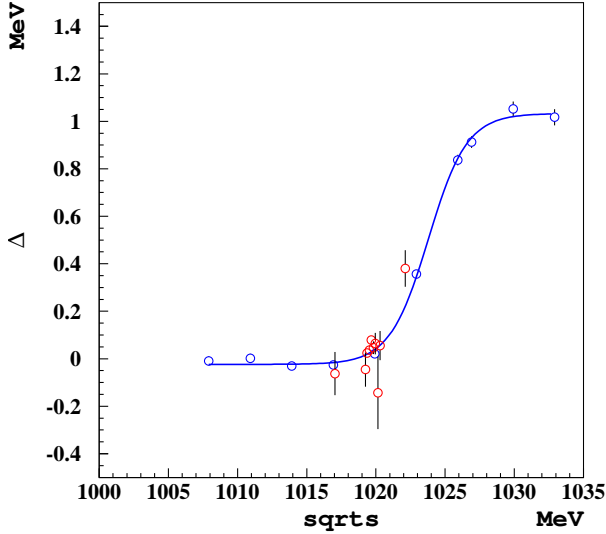}
\caption{correction to the $\eta$ mass evaluated by MC as a function of $\sqrt{s}$. Points without error bars: MC; continuous line: fitted curve to the MC points; points with error bars: DATA.} \label{fig:isrcorr}
\end{figure}

\begin{figure}
\includegraphics[width=0.5\textwidth]{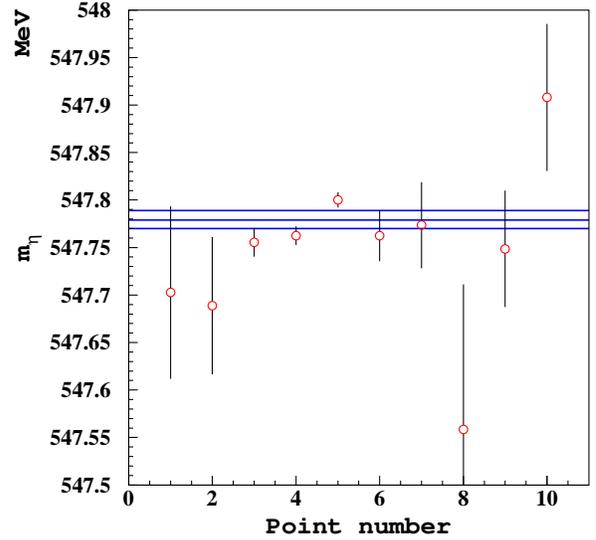}
\caption{$m_{\eta}$ as a function of $\sqrt{s}$ after the ISR correction. The middle line is the mean value while the two line above and below are $+/-$ 1 $\sigma$ systematic.} \label{fig:metacorr}
\end{figure}

The absolute energy scale of the calorimeter and the linearity of the energy response was checked using the
$e^+e^- \to e^+ e^- \gamma$ events and the $\pi^+ \pi^- \gamma$ events. The energy of the $\gamma$ can be determined using the
two charged tracks in the Drift Chamber and then compared to the reconstructed cluster energy. A linearity better than 2 \% was found while
the absolute scale was found to be calibrated at better than 1 \%. These systematic uncertainties result in just 4 keV for the scale
and 4 keV for the linearity on the value of the reconstructed mass. The measurement shows  very small sensitivity to the calorimeter calibration because, as explained before, the kinematic fit overconstraints the photon energy with the cluster positions. \\
For this reason it is important to evaluate the systematic error due to the misalignment of single modules in the calorimeter. This has been done evaluating the value of the mass as a functon of the position of the photons in the calorimeter. A spread of about 10-15 keV was found and assumed as systematic error. \\
Systematics due to the particular choice of the cut on the Dalitz plot shown in fig.\ref{fig:dalitz} was also determined to be 12 keV, while the cut on the $\chi^2$ pratically doesn't have any effect on the value of the $\eta$ mass (0.7 keV). \\
The measured value of the mass is, instead, very sensitive to the energy in the center of mass of the $\eta \gamma$ system. Due to initial state radiation emission (ISR) the available center of mass energy  is a bit lower than
the $\sqrt{s}$ of the $e^+ e^-$  beams measured using $e^+ e^- \to e^+ e^-$ events. A variation of 100 keV of the measured mass value is predicted by the MC simulation. Since this correction is realtively large we have checked the simulation of ISR emission in the MC in the following way:
\begin{itemize}
\item the 
correction to apply to the fitted value in order to obtain the real value of the mass has been determined as a function of $\sqrt{s}$ and
shown in fig.\ref{fig:isrcorr};
\item  the whole data taking has been devided in ranges of $\sqrt{s}$ using 8 points for the on peak data and  two off-peak points at 1017 MeV and 1022 MeV;
\item  the value of the mass obtained for each value of $\sqrt{s}$ has been corrected according the MC prediction;
\item  the residual spread of these points has been taken as systematic error (8 keV) (see fig. \ref{fig:metacorr}). 
\end{itemize}
 
All these studies have been done also for the $\pi^0$ mass using  the $\phi \to \pi^0 \gamma$ events. The ratio of the two masses $r=m_{\eta}/m_{\pi^0}$ has also been studied. All the  contributions to the systematic error are summarized in table \ref{tab:syst}.
\begin{table*}
\begin{center}
\begin{tabular}{|c|c|c|c|}
\hline \hline
systematic effect & $m_{\eta}$ (keV) & $m_{\pi^0}$ (keV) & $m_{\eta}/m_{\pi^0} \times 10^{-5}$ \\
\hline
Calorimeter energy scale & 4 & 1 & 5.6 \\
\hline
Calorimeter not linearity & 4 & 11 & 31 \\
\hline
Vertex position & 4 & 6 & 19\\
\hline
Angular uniformity $\phi$ & 15 & 12 & 37 \\
\hline
Angular uniformity $\theta$ & 10 & 44 & 120\\
\hline
ISR effect & 8  & 9 & 28 \\
\hline
Dalitz plot cut slope & 12 & 4 & 15 \\
\hline
Dalitz plot cut constant & 12 & 1.9 & 10 \\
\hline
$\chi^2$ cut & 0.7 & 4 & 13 \\
\hline
\textbf{overall} & 27 & 49 & 136 \\
\hline
\end{tabular}
\end{center}
\caption{Summary of all evaluated systematic effects. The Dalitz plot slope and constant refers to the change of  
slope or  intercept of the stright line cut in fig. \ref{fig:dalitz}.} \label{tab:syst}
\end{table*}

\begin{figure}
\includegraphics[width=0.4\textwidth]{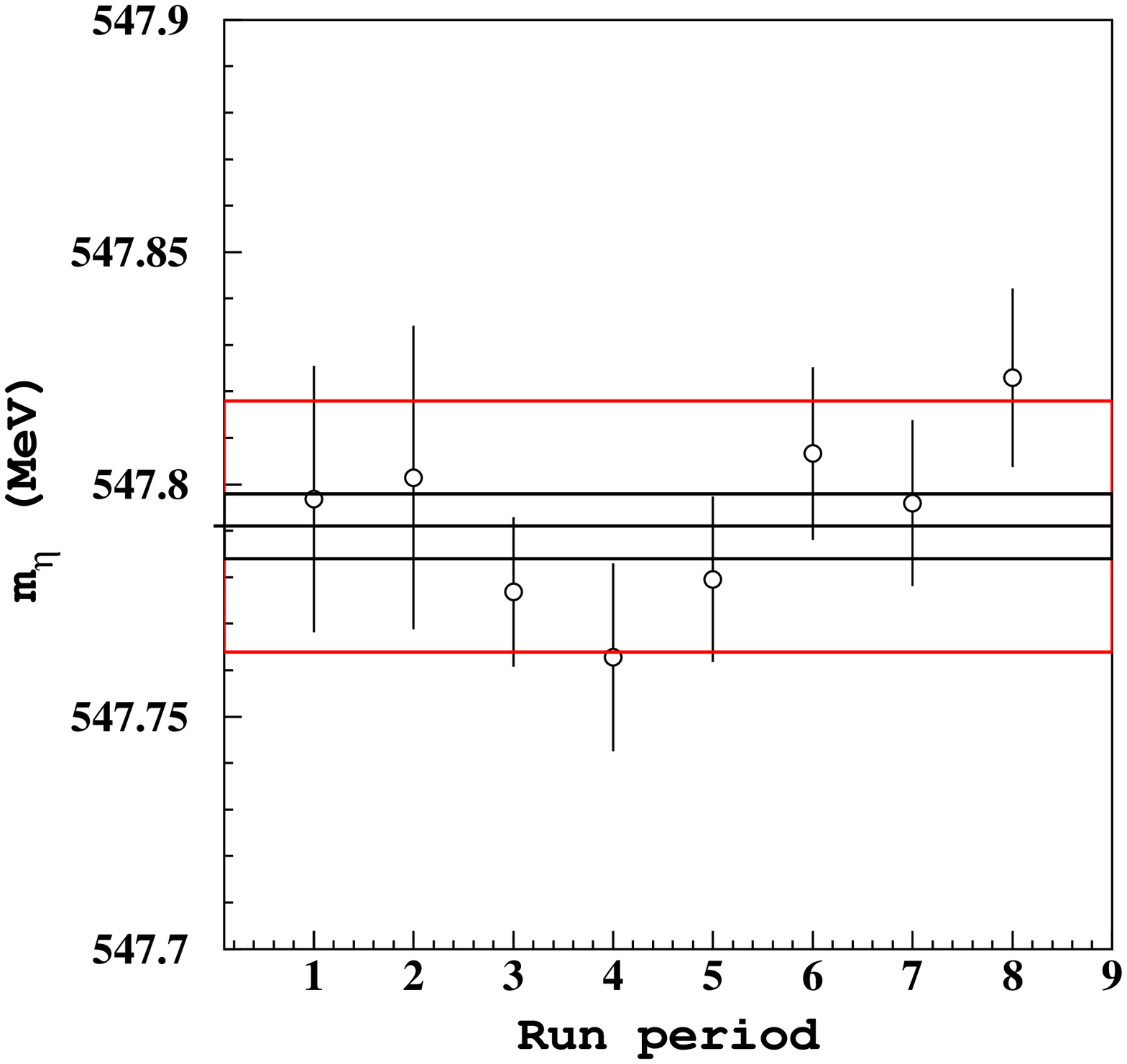}
\includegraphics[width=0.4\textwidth]{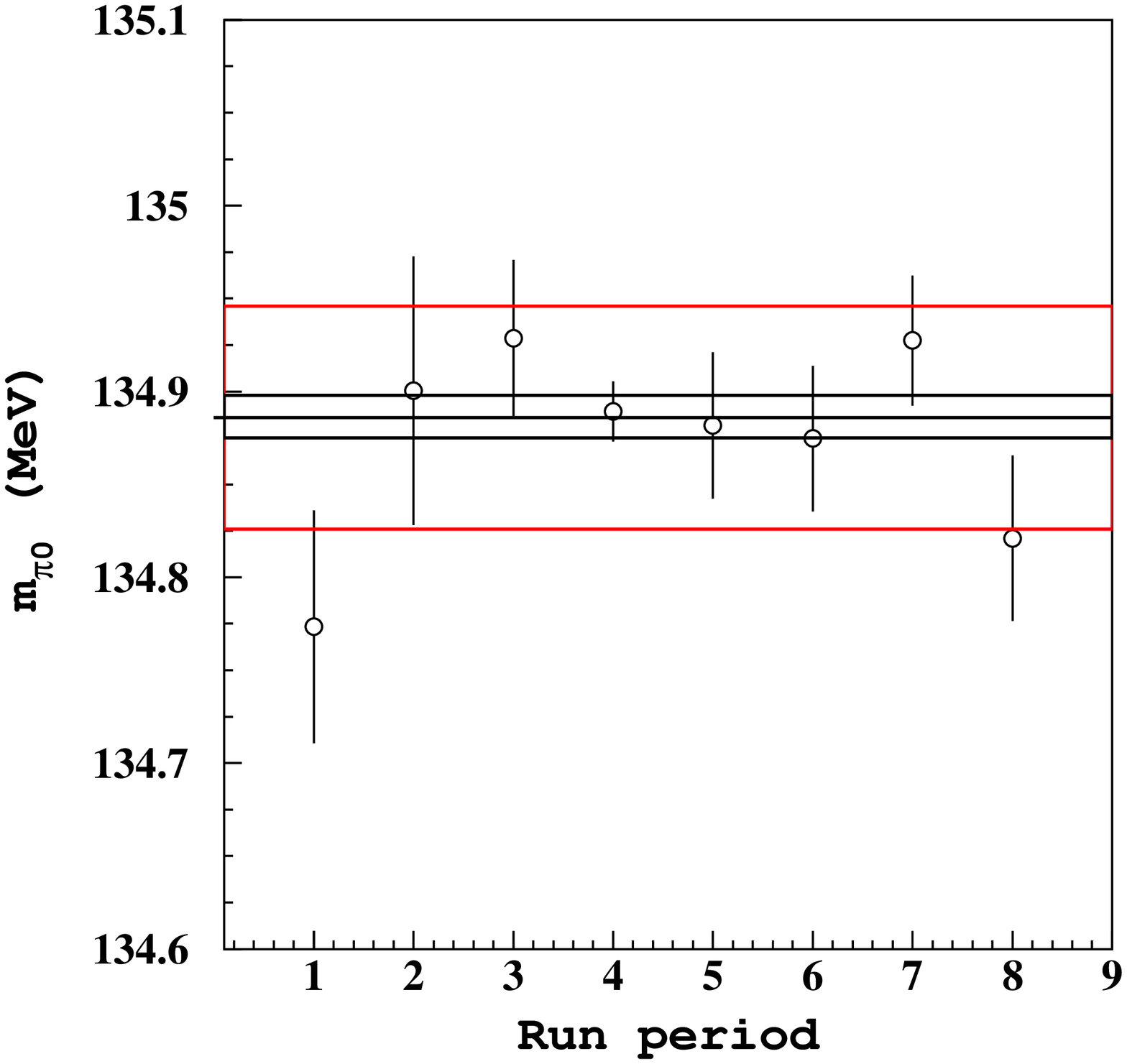}
\includegraphics[width=0.4\textwidth]{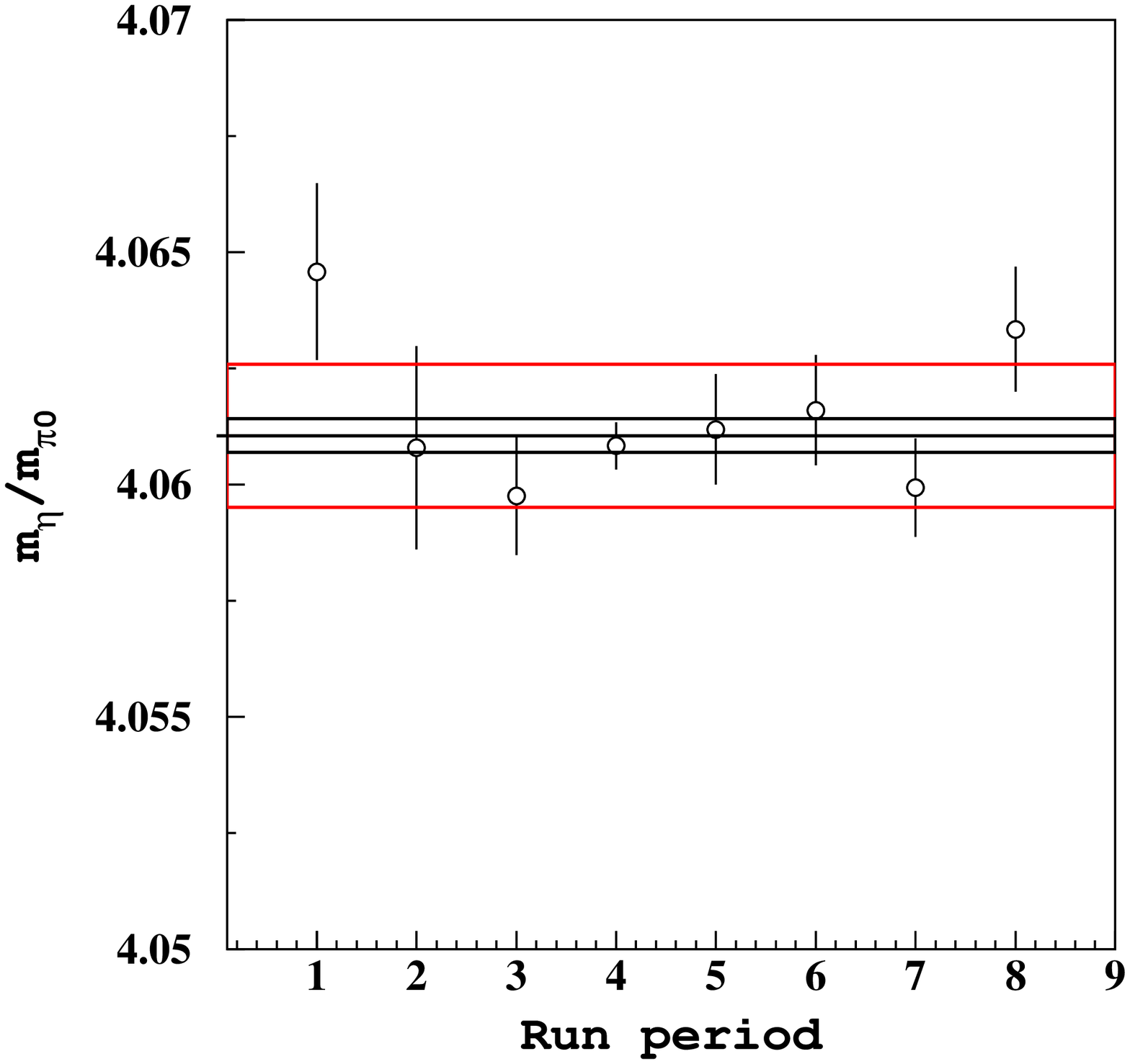}
\caption{From top to bottom: values of the $m_{\eta}$, $m_{\pi^0}$ and $r$ as a function of the data acquisition periods.The mean line is the fitted value to a constant, the small rectange the statistical error and the large rectangle the systematic error.} \label{fig:valuevsruns}\end{figure}

Finally stability versus run conditions are checked dividing the whole period of DATA taking 2001-2002 in 8 sub-periods and determining the 
value of $m_{\pi^0}$, $m_{\eta}$ and $m_{\eta}/m_{\pi^0}$ as a function of the acquisition period. The results are shown in fig. \ref{fig:valuevsruns}. \\
The values shown in the figure have been fitted with a constant and the results of the fits are shown in tab. \ref{tab:fitfinal}.
\begin{table}
\begin{center}
\begin{tabular}{|c|c|c|c|c|}
\hline
\multicolumn{5}{|c|}{Fit results} \\
\hline
&Value & Error& $\chi^2/n.d.f$ & C.L \\
\hline
$m_{\eta}$ & 547791  keV &  7 keV  & 6.9/7 & 45\%\\
\hline
$m_{\pi^0}$ & 134886 keV  & 12 keV  & 7.7/7 & 34\% \\
\hline
$m_{\eta}/m_{\pi}$ & 4.0610 & 0.0004 & 8.9/7 & 26\% \\
\hline
\end{tabular}
\end{center}
\caption{Fit results to the 8 data points.} \label{tab:fitfinal}
\end{table}
\section{Computation of the final result.}
We have two different ways to extract the value of the $\eta$ mass.  We can use the ratio $m_{\eta}/m_{\pi^0}$ obtaining:
\[
\frac{m_{\eta}}{m_{\pi^0}} = 4.0610 \pm 0.0004 \mathrm{(stat.)} \pm 0.0014 \mathrm{(syst.)}
\]
from which using the PDG2006 \cite{PDG} value of the $\pi^0$ mass ($m_{\pi^0}$ = 134976.6 $\pm$ 0.6 \, keV) we obtain:
$m_{\eta} =  548140 \pm 50_\mathrm{(stat.)} \pm 190_\mathrm{(syst.)}$ keV.  \\
Alternatively we can use directly the value of $m_{\eta}$ coming from the fit. For this purpose we need to calibrate
the $\sqrt{s}$ with high precision ($m_{\eta}/m_{\pi^0}$ is in fact $\sqrt{s}$ independent while the two values of $m_{\eta}$ and
$m_{\pi^0}$ are fully correlated with the $\sqrt{s}$ measuremnt). The absolute $\sqrt{s}$ scale has been calibrated using the $m_{\phi}$ value measured by CMD-2 \cite{PhiMass}. To this pourpose the cross section $e^+ e^- \to \phi \to K_S K_L$ has been measured as a function of the $\sqrt{s}$ using the two off-peak points at $\sqrt{s} = 1017 MeV$ and $\sqrt{s} = 1022$ MeV together with the on-peak data. The $\phi$ resonance curve has been fitted using the CMD-2 parametrization that takes in account ISR and threshold effect in $K_S K_L$ production. The central value of the $\phi$ mass has been measured obtaining $m_{\phi} = 1019.329 \pm 0.011$ MeV \cite{Antonelli}. The difference respect to the CMD-2 value $m_{\phi \,\, CMD-2} = 1019.483 \pm 0.011_{stat.} \pm 0.025_{syst.}$ sets our absolute $\sqrt{s}$ calibration.

This means that our measurements can be regarded as  a measurement of the ratio $m_{\eta}/m_{\phi CMD2}$ and $m_{\pi^0}/m_{\phi CMD2}$:
\[
\begin{array}{rl}
m_{\eta}/m_{\phi CMD2}&= 0.537403 \pm 0.000007 \, \mathrm{(stat.)}  \\
&\pm 0.000026 \, \mathrm{(syst.)} \pm 0.000006 \, (m_{\phi} \,\mathrm{stat.})
\end{array}
\]
and
\[
\begin{array}{rl}
m_{\pi^0}/m_{\phi CMD2}&= 0.132328 \pm 0.000012 \, \mathrm{(stat.)} \\
&\pm 0.000048 \, \mathrm{(syst.)} \pm 0.000001  \, (m_{\phi} \, \mathrm{stat.})
\end{array}
\]
We can extract the $\pi^0$ and $\eta$ mass using the value of CMD-2 $m_{\phi}$ obtaining:

\begin{eqnarray*}
m_{\pi_0} & = & 134.906 \pm 0.012 \mathrm{(stat.)} \pm 0.048 \mathrm{(syst.)} \quad \mathrm{MeV} \\
m_{\eta} & = & 547.873 \pm 0.007 \mathrm{(stat.)} \pm 0.031 \mathrm{(syst.)} \quad \mathrm{MeV}
\end{eqnarray*}
where the statistical and sytematic error have been  independently summed in quadrature.

The $\pi^0$ mass is in agreement  at 1.4 $\sigma$ level with the PDG06 value, validating the whole procedure. 

\begin{figure}[htbp!]
\begin{center}
\includegraphics[width=0.45\textwidth]{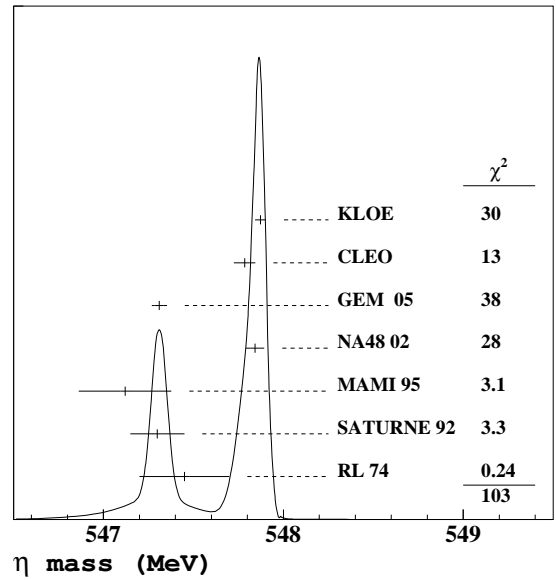}
\end{center}
\caption{$\eta$ mass measurements, see text for the references. The continuous line has been computed using the PDG procedure \cite{PDG}, pag. 14.} \label{fig:ideofinal}
\end{figure}

The value of the $\eta$ mass obtained here has been compared with the previous measurements in fig. \ref{fig:ideofinal}. This result confirms the measured value of the NA48 at $0.6$$\sigma$, with an error reduced by a factor 2 and is 11$\sigma$ away from the GEM result, it is  also compatible, at $1.4 \sigma$ level, with less accurate CLEO result \cite{CLEO}.  

\end{document}